\documentclass[aps,pre,reprint,showpacs,showkeys]{revtex4-1}
\usepackage{amsfonts,amssymb,amsmath}
\usepackage{bm,bbm,euscript,mathrsfs}
\usepackage[usenames]{color}
\usepackage[utf8]{inputenc}
\usepackage[T1]{fontenc}
\usepackage{textcomp}
\usepackage{graphicx}
\usepackage[tight,sf]{subfigure}
\usepackage{hyperref}
\newcommand{\arcsinh}{\mathrm{arcsinh}}

\begin{document}

\title{Constrained metric variations and emergent equilibrium surfaces}

\author{\firstname{Jemal} \surname{Guven}}
\email[]{jemal@nucleares.unam.mx}
\affiliation{Instituto de Ciencias Nucleares, Universidad Nacional Aut\'onoma de M\'exico\\
Apdo. Postal 70-543, 04510 M\'exico, DF, MEXICO}

\author{\firstname{Pablo} \surname{V\'azquez-Montejo}}
\email[]{pvazquez@correo.cua.uam.mx}
\affiliation{Departamento de Matem\'aticas Aplicadas y Sistemas,\\
Universidad Aut\'onoma Metropolitana-Cuajimalpa, C.P. 01120, M\'exico D.F., MEXICO}


\begin{abstract}
Any surface is completely characterized by a metric and a symmetric tensor satisfying the Gauss-Codazzi-Mainardi equations (GCM), which identifies the latter as its curvature.  We demonstrate that physical questions relating to a surface described by any Hamiltonian involving only surface degrees of freedom can be phrased completely in terms of these tensors without explicit reference to the ambient space: the surface is an emergent entity. Lagrange multipliers are introduced to impose GCM as constraints on these variables and equations describing stationary surface states derived. The behavior of these multipliers is explored for minimal surfaces, showing how their singularities correlate with surface instabilities.
\end{abstract}


\maketitle

\section{Introduction}

Surfaces occur as approximations of physical systems at almost all energy scales \cite{Nelson}. More often than one would expect the only relevant degrees of freedom are the ones associated with the geometric
configuration of the surface itself and its behavior is described completely by a Hamiltonian or an action
constructed using the geometrical  invariants of this surface.  This may be something as simple as
the area--representing the energy of an interface or a soap film \cite{Soap}--or its relativistic
analogue which represents the area of the worldsheet swept out in the course of the evolution of a
string,  be it a fundamental extended object or--more conservatively--some effective description of one\cite{Polyakov, Strings}.\\
Typically the Hamiltonian defined on a surface, $\Gamma: \{u^1,u^2\} \rightarrow
\mathbf{X}(u^1,u^2)$, is constructed by forming suitable scalars using the induced metric $g_{ab}$,
the curvatures $K_{ab}$ and their covariant derivatives:
\begin{equation} \label{Hdef}
H = \int dA\, {\cal H}[g_{ab}, K_{ab}]\,;
\end{equation}
for simplicity, we consider only surfaces embedded in three-dimensional Euclidean space,
$\mathbb{E}^{3}$. The important point is that the functions $\mathbf{X}$ tend not to  appear
explicitly in $H$. Surface area with ${\cal H}=1$, depending on the metric through its determinant,
$d A= d^2 u\,\sqrt{g}$, is the simplest example. If the tensors $g_{ab}$ and $K_{ab}$ in Eq.
(\ref{Hdef}) are to represent a surface, however,  they will need to be consistent with the
Gauss-Codazzi (GC) and Codazzi-Mainardi (CM) equations,
\begin{equation} \label{GCM2D}
{\cal G} =0\,,\quad {\cal C}_a =0\,,
\end{equation}
where
\begin{subequations}
\begin{eqnarray}
{\cal G} &:=& {\cal R} - K^2 + K_{ab} K^{ab} \,;\label{GC}\\
{\cal C}_a &:=& \nabla^{b}\left( K_{ab} - g_{ab} K\right) \,, \label{CM}
\end{eqnarray}
\end{subequations}
which occur as integrability conditions on the structure equations defining how the unit tangents and normals rotate as one moves along the surface. Here $\nabla_a$ is the covariant derivative compatible with $g_{ab}$; ${\cal R}$ is the corresponding Ricci scalar curvature and $K$ represents the trace of $K^{a}_{\phantom{a}b}$, $K = g^{ab}K_{ab}$. Conversely, one of the corner pieces of nineteenth century geometry is the assertion that any two tensor fields, $g_{ab}$ and $K_{ab}$, satisfying Eqs. (\ref{GCM2D}) will represent some surface $\mathbf{X}$, with induced metric $g_{ab}$ and extrinsic curvature $K_{ab}$, unique up to Euclidean motions \cite{Spivak}. This will also be crucial. Indeed, even if $H$ depended only on the metric, this metric knows there is an extrinsic curvature tagging along.\\
In this Letter, we will show that it is always possible to rephrase the variational properties of
surfaces in terms of a theory {\it of  gravity} involving a metric, coupled to an auxiliary field
$K_{ab}$, without any explicit reference to the embedding functions themselves: the surface itself
is an emergent entity. In this framework Eqs. (\ref{GCM2D}) are enforced by introducing Lagrange
multipliers, which permits one to treat these two tensors as independent variables.\\
This approach contrasts dramatically with the familiar approach in terms of harmonic maps \cite{Polyakov, Eels}, or its natural extension--when curvatures are involved--in terms of auxiliary variables \cite{auxil}: here, the surface does not materialize until these constraints are applied. A comparison between this framework and the latter is presented in Appendix \ref{appauxil}. Relevant antecedents motivating this work can be found in Barbour, Foster and Ó Murchadha's ``Relativity without relativity'' \cite{Murchadha}, Sorkin's treatment of field theory in Minkowski space \cite{Sorkin}, or Lomholt and Miao's discussion of the ambiguities associated with the GCM equations \cite{Lomholt}. It also shares features with the framework, developed in \cite{Paperfold} in the context of paper folding, for enforcing local geometrical constraints.\\
A peculiarity of two-dimensional surfaces is that the multipliers assemble into a spatial vector field.  If $H$ depends only on the intrinsic geometry, this vector field can be identified in equilibrium as a generator of surface isometries; if $H$ depends also on the curvature $K_{ab}$, on the other hand, it is identified with a conformal transformation. The role of the multipliers themselves, however, is not  to displace the surface.  This identification is a two-dimensional accident: they represent the strength of the interaction coupling the tensor field $K_{ab}$ to the metric on the Riemannian manifold in the formation of the equilibrium surface. The surface Euler-Lagrange equations are derived by examining the flows generated by this vector field. Its behavior will be explored in detail for area minimizing surfaces. In the case of a catenoid bridging two rings, the relevant isometry will be identified explicitly, and the connection between the singularities in this vector field and the presence of instabilities emphasized. This framework appears to provide a new approach to analyzing the instability of equilibrium surfaces.

\section{Surface variational principles without surfaces}

Consider the following effective action or energy
\begin{eqnarray} \label{HC}
H_C [g_{ab},K_{ab},\Lambda,\lambda^{a}] & = & H[g_{ab},K_{ab}] \, \nonumber \\
&+& I[g_{ab},K_{ab},\Lambda,\lambda^{a}] \,,
\end{eqnarray}
where
\begin{equation}
I = \frac{1}{4}\int dA \, \Lambda\, {\cal G} + \frac{1}{2}\int dA \, \lambda^{a} {\cal C}_a \,.
\end{equation}
The Lagrange multipliers fields $\Lambda$ and $\lambda^{a}$ enforce the GC and CM equations, Eqs. (\ref{GCM2D}), as constraints on the variables $g_{ab}$ and $K_{ab}$. In Eq. (\ref{HC}) one is now free to treat $g_{ab}$ and $K_{ab}$ as independent variables. The variation of $H_C$ is given by
\begin{eqnarray} \label{deltaHcfull}
\delta H_C &=& \int dA\, \left( -\frac{1}{2} (T^{ab} + {\cal T}^{ab}) \, \delta g_{ab}
+ (H^{ab} +  {\cal H}^{ab})  \, \delta K_{ab} \right)\nonumber \\
&+ & \int dA \,\nabla_a Q^{a}\,,
\end{eqnarray}
where the manifestly symmetric second rank tensors $T^{ab}$ and $H^{ab}$, are associated with the
variation of $H$ with respect to $g_{ab}$ and $K_{ab}$;
${\cal T}^{ab}$  and ${\cal H}^{ab}$ are their counterparts for the constraining term $I$:
\begin{subequations} \label{TH}
\begin{align}
T^{ab}&=-2\frac{\delta H}{\delta g_{ab}} \,, &\quad {\cal T}^{ab} &= -2 \frac{\delta I}{\delta g_{ab}}\,;\\
 H^{ab} &= \frac{\delta H}{\delta K_{ab}} \,, &\quad {\cal H}^{ab} &= \frac{\delta I}{\delta K_{ab}}\,.
\end{align}
\end{subequations}
In Eq. (\ref{deltaHcfull}) $Q^a$ represents all of the terms that have been collected in a
divergence after integration by parts.
\vskip1pc \noindent
The Euler-Lagrange equations for $g_{ab}$ and $K_{ab}$ describing the equilibrium states of the
surface are given respectively by:
\begin{subequations} \label{THcab}
\begin{eqnarray}
T^{ab} + {\cal T}^{ab} &=& 0\,; \label{Tcab}\\
H^{ab} + {\cal H}^{ab} &=& 0\,, \label{Hcab}
\end{eqnarray}
\end{subequations}
supplemented with Eqs. (\ref{GCM2D}). Eqs. (\ref{THcab}) are analogues of the Einstein equations in general relativity. The technicalities of the variations with respect to $g_{ab}$ and $K_{ab}$ are themselves
straightforward (see, for example \cite{MTWetc}). One identifies
\begin{subequations}
\begin{eqnarray}
{\cal T}^{ab} &=& -\frac{1}{2} \left( \nabla^a \nabla^b - g^{ab} \nabla^2 + {\cal R}^{ab}\right) \Lambda \nonumber \\
&+& \frac{1}{2} \left[\nabla^a (\lambda^{c} K^b_{\phantom{b}c}) + \nabla^b (\lambda^{c} K^a_{\phantom{a}c}) \right] \nonumber \\
&-& \frac{1}{2} \nabla_c \left[\lambda^{c} K^{ab} + g^{ab} \lambda^d K^c_{\phantom{c}d}\right] \,; \label{Tab}\\
{\cal H}^{ab} &=& \frac{1}{2} \left(K^{ab} -g^{ab} K \right) \, \Lambda \nonumber \\
&+& \frac{1}{4}\left( \nabla^a \lambda^b + \nabla^b \lambda^a\right) - \frac{1}{2}\nabla_c \lambda^{c} g^{ab}\,. \label{Hab}
\end{eqnarray}
\end{subequations}
The task is now to solve, if only implicitly,  Eqs. (\ref{THcab})  for the multipliers. This is
facilitated by organizing the two tensors ${\cal T}^{ab}$ and ${\cal H}^{ab}$ in a more
geometrically transparent way.\\
Introduce the {\it Lie} derivative along the vector field $\lambda^a$ on the Riemannian manifold,
which acts on the tensors $g_{ab}$ and $K_{ab}$ as follows
\begin{subequations}
\begin{eqnarray}
{\cal L}_{\lambda} g_{ab} &=& \nabla_a \lambda_b + \nabla_b \lambda_a \,;
\label{deltalgab}\\
{\cal L}_{\bf \lambda}  K_{ab} &=& \left (\nabla_c K_{ab} + K_{ac}\nabla_b  + K_{bc}\nabla_a\right) \lambda^c\,,
\label{deltalKab}
\end{eqnarray}
\end{subequations}
and define analogues ${\cal L}_\Lambda$ for the scalar $\Lambda$
\begin{subequations}
\begin{eqnarray}
{\cal L}_{\Lambda} g_{ab} &=& 2 K_{ab} \Lambda \,; \label{deltaLgab}\\
{\cal L}_{\Lambda}  K_{ab} &=& \left(- \nabla_a \nabla_b + K_{ac} K^c{}_b \right) \Lambda \,.
\label{deltaLKab}
\end{eqnarray}
\end{subequations}
To motivate these definitions, consider for a moment a surface ${\bf X}$ with tangent vectors $\mathbf{e}_a = \partial_a \mathbf{X}$ and unit normal vector $\mathbf{n}$. One can then construct a space vector ${\bf \Lambda}= \lambda^a \mathbf{e}_a + \Lambda\mathbf{n}$, with tangential components $\lambda^a$ and normal component $\Lambda$. Now define
\begin{subequations} \label{deltagKab}
\begin{eqnarray}
{\cal L}_{\bf \Lambda} g_{ab} &=& {\cal L}_{\Lambda} g_{ab} + {\cal L}_{\lambda} g_{ab}
\,; \label{deltagab}\\
{\cal L}_{\bf \Lambda}  K_{ab} &=& {\cal L}_{\Lambda} K_{ab} + {\cal L}_{\lambda} K_{ab}
\,. \label{deltaKab}
\end{eqnarray}
\end{subequations}
The induced metric $g_{ab} = \mathbf{e}_a\cdot \mathbf{e}_b$, and the extrinsic curvature tensor
$K_{ab}=\mathbf{e}_a\cdot \nabla_b \mathbf{n}$, then transform respectively by Eqs.
(\ref{deltagKab}) under the flow generated by the vector field ${\bf \Lambda}$ (see, for example,
\cite{Defos}).\\
It should be stressed that neither of the definitions Eqs. (\ref{deltagKab}) make any reference to
the embedding functions $\mathbf{X}$, the  identifications of $g_{ab}$ and $K_{ab}$ in terms of these functions,
or the assembly of $\Lambda$ and $\lambda^a$ into a space vector;  more importantly, despite the
shorthand, it is not even appropriate to think of ${\bf \Lambda}$ as a spatial vector field. If this
were the case the flow defined by Eqs. (\ref{deltagKab}) would displace the surface geometry away
from equilibrium.  It is, in fact, a two-dimensional accident that a space vector can be constructed
using the multipliers $\lambda^a$ and $\Lambda$. In this context, the role played by ${\bf \Lambda}$
contrasts with the one played by the lapse and shifts in the Hamiltonian formulation of general
relativity where the analogs of the  GCM constraints for a spatial hypersurface embedded in a
Riemannian manifold are the generators of normal  and tangential deformations of this hypersurface
\cite{ADM}. Their role here is not to displace the surface: rather they are the generalized forces
coupling the tensor fields $g_{ab}$ and $K_{ab}$ to form the induced metric and extrinsic curvature
of the surface.\\
The motivation  for introducing Eqs. (\ref{deltagKab}) is that it is now possible to cast the
tensors ${\cal T}^{ab}$ and ${\cal H}^{ab}$ in the remarkably simple form
\begin{subequations}
\begin{eqnarray}
{\cal T}^{ab} &=& \frac{1}{4} \left(g^{ab} K^{cd} -g^{cd} K^{ab} \right) \, {\cal L}_{\bf  \Lambda} g_{cd} \nonumber\\
&+& \frac{1}{2} \, \left(g^{ac} g^{bd} - g^{ab} g^{cd} \right) \, {\cal L}_{\bf  \Lambda} K_{cd}\,; \label{TabdelgabKab}\\
{\cal H}^{ab} &=& \frac{1}{4} \left(g^{ac} g^{bd}- g^{ab} g^{cd} \right) {\cal L}_{\bf  \Lambda} g_{cd}\,, \label{HabnoT}
\end{eqnarray}
\end{subequations}
linear in ${\cal L}_{\bf  \Lambda} g_{ab}$ and ${\cal L}_{\bf  \Lambda} K_{ab}$.\\
{\bf A useful identity:} Using the intrinsic definition of ${\cal R}$, and its extrinsic
counterpart implied by  the GC equation, ${\cal C}_\perp=0$, one identifies two
equivalent expressions for ${\cal L}_{\bf \Lambda} {\cal R}$:
\begin{subequations}
 \begin{eqnarray} \label{delRGCM}
 {\cal L}_{\bf \Lambda} {\cal R} &=& -2 \left({ \cal R}^{ab}  {\cal L}_{\bf  \Lambda}
g_{ab}+(K^{ab}-K g^{ab}){\cal L}_{\bf  \Lambda} K_{ab}\right) \\
&=&  \left(\nabla^a \nabla^b - g^{ab} \nabla^2 - {\cal R}^{ab}\right) {\cal L}_{\bf  \Lambda}
g_{ab}\,.
\end{eqnarray}
\end{subequations}
As a consequence, the projection of ${\cal T}^{ab}$, given by Eq. (\ref{TabdelgabKab}), on $K_{ab}$ can be
cast completely in terms of ${\cal L}_{\Lambda} g_{ab}$:
\begin{equation} \label{KabTab}
K_{ab} {\cal T}^{ab} = \frac{1}{4} \big(g^{ab} \nabla^2 - \nabla^a \nabla^b +K^{ac}
K_c^{\phantom{c}b}-g^{ab} K_{cd} K^{cd}\big) \, {\cal L}_{\bf  \Lambda} g_{ab}\,.
\end{equation}
The significance of this identity will soon be apparent.
\section{Two cases of interest: tension and bending}

{\bf  Gravitational Impostors:} Let us first consider a Hamiltonian depending only on the metric, so that ${\cal H}= {\cal H}[g_{ab}]$ in Eq. (\ref{Hdef}).\\
In this case $ H^{ab}=0$ so that Eq. (\ref{Hcab}) implies that ${\cal H}^{ab}=0$. The identity
(\ref{HabnoT}) in turn implies that the vector ${\bf \Lambda}$, treated as a space vector,
can be  identified as the generator of an isometry, in the sense that ${\cal L}_{\bf  \Lambda} g_{cd}=0$.
The identity (\ref{KabTab}) then implies that $K_{ab} {\cal T }^{ab}=0$.  As an immediate
consequence of Eq. (\ref{Tcab}), the Euler-Lagrange equation $- K_{ab} \, T^{ab} = 0$ follows:  a
surprisingly short story once the role of ${\bf \Lambda}$ as generator of isometries is recognized.
Notice that  $T^{ab}$ generally does not vanish; thus ${\cal T}^{ab}\ne 0$. Eq. (\ref{TabdelgabKab})
then implies that the isometry is non-trivial; for if ${\cal L}_{\bf  \Lambda} K_{ab} \ne 0$,  ${\bf
\Lambda}$ cannot be a Euclidean motion.\\
In particular, in  the case ${\cal H}$ is some constant $\sigma$, so that $H$ is proportional to
area, one identifies $ T^{ab}=- \sigma g^{ab}$, and the Euler-Lagrange equation reduces to
$K=0$.  A familiar statement is recovered:  the stationary states are minimal surfaces.\\
{\bf Bending energy:} A less simple example is provided by the Polyakov or Helfrich bending energy, quadratic in
curvature, with ${\cal H}[g_{ab}, K_{ab}] =K^2/2$.  It is the simplest non-topological conformal
invariant of an embedded two-dimensional surface \cite{Willmore}; it also provides an
extraordinarily robust mesoscopic description of fluid membranes \cite{HelCan}. Now  $ T^{ab} = K (2
K^{ab} - 1/2 g^{ab} K)$ and $ H^{ab}= K g^{ab}$;  as a result of the latter, Eq. (\ref{HabnoT})
reads ${\cal L}_{\bf  \Lambda} g_{ab} = 4 K g_{ab}$. Thus ${\bf \Lambda}$ generates a conformal
transformation, scaling locally with the mean curvature.  Eq. (\ref{KabTab}) then implies that
$K_{ab} {\cal T}^{ab}=(\nabla^2 -K_{ab}K^{ab}) K$ so that the Euler-Lagrange shape equation is given
by
\begin{equation} \label{elK2}
\left(-\nabla^2 +{\cal R}-\frac{1}{2} K^2\right)K = 0\,,
\end{equation}
a surprisingly pithy derivation that compares favorably with any of its better established counterparts, all the more so because this framework  was not developed to compete on this level.\\
The linearity of the Euler-Lagrange equations in the multipliers permits one to treat more
complicated energies, the Helfrich Hamiltonian ${\cal H}= (K-K_0)^2/2 + \sigma $, with spontaneous
curvature $K_0$ and constrained area, for example.

\section{Multipliers and instabilities for minimal surfaces}

\begin{figure}[htb]
\begin{center}
\begin{tabular}{cc}
\subfigure[]{\includegraphics[scale=0.075]{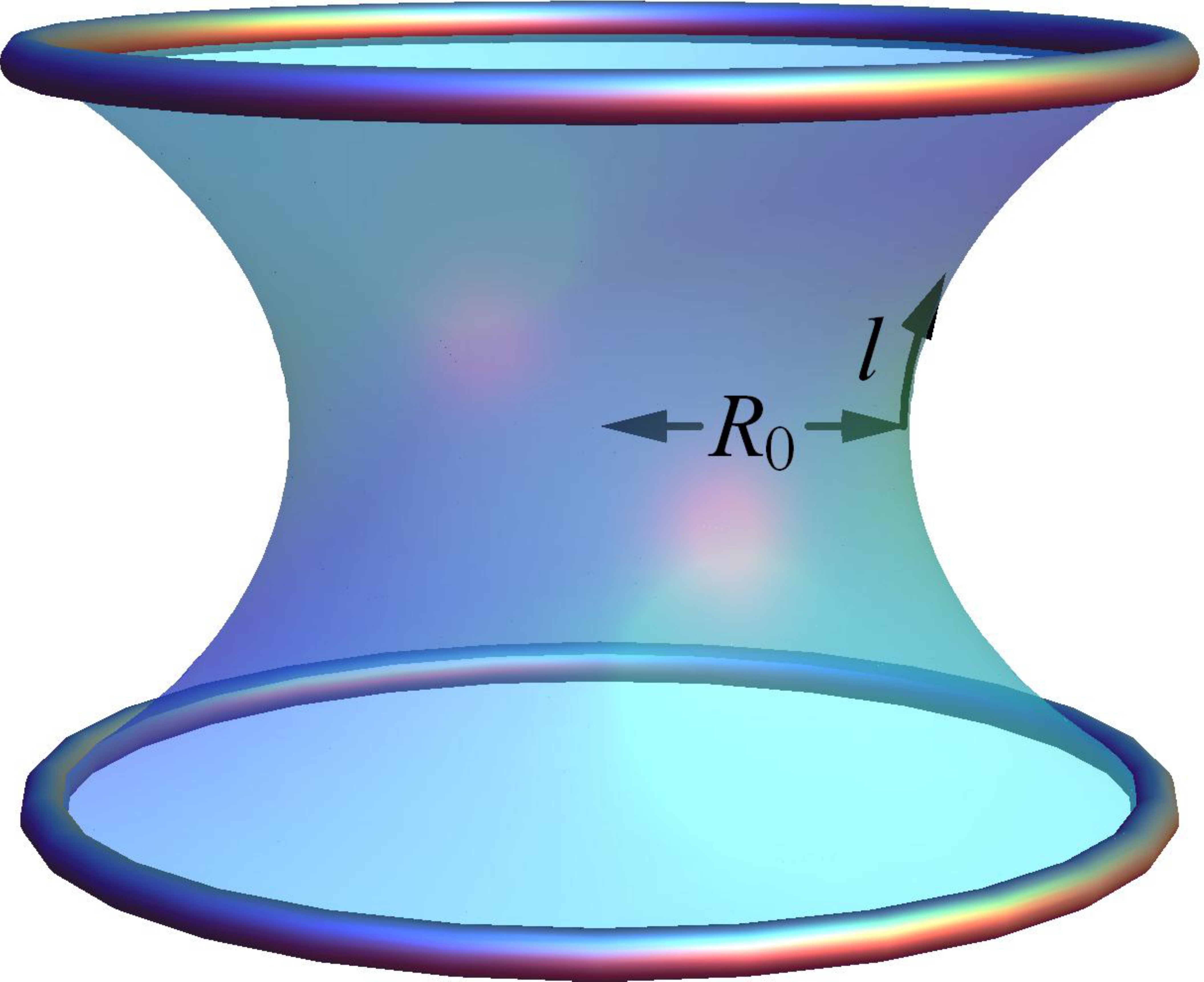}} &
\subfigure[]{\includegraphics[scale=0.33]{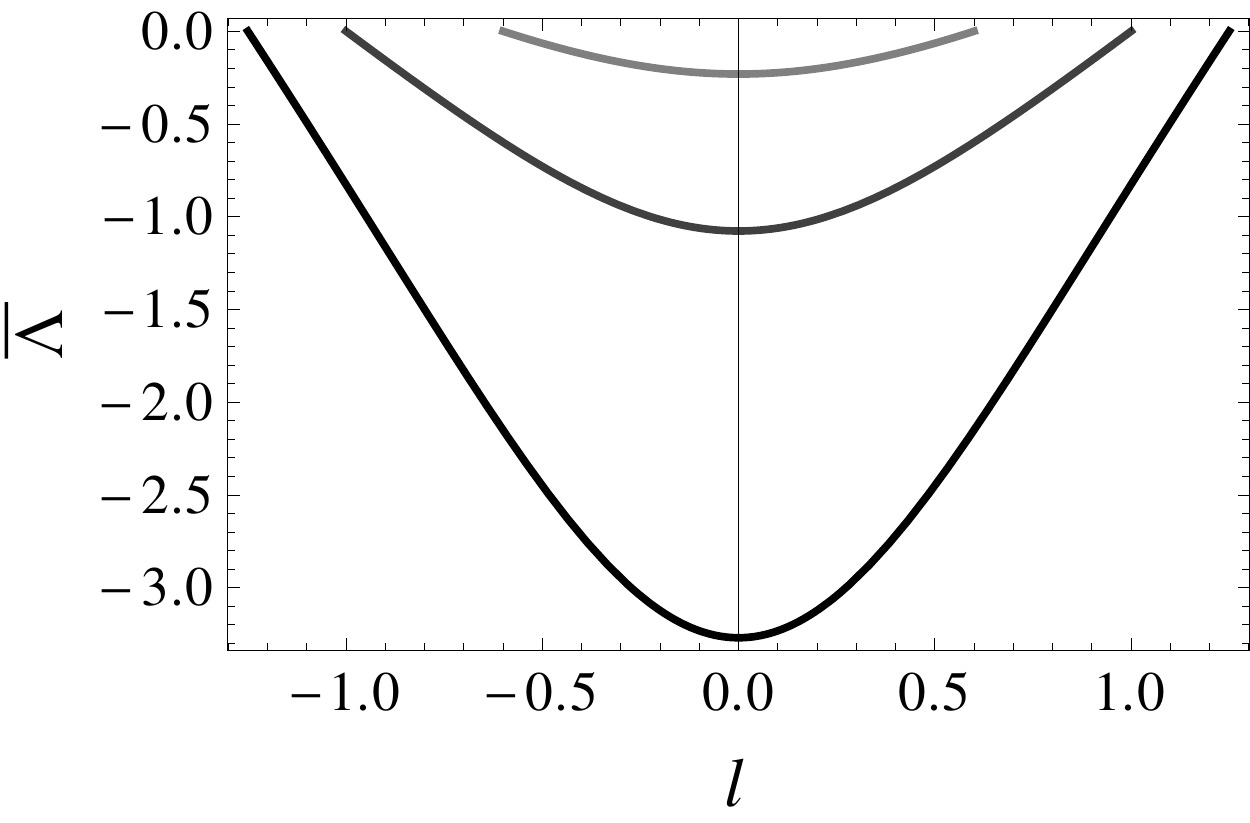}} \\
\subfigure[]{\includegraphics[scale=0.33]{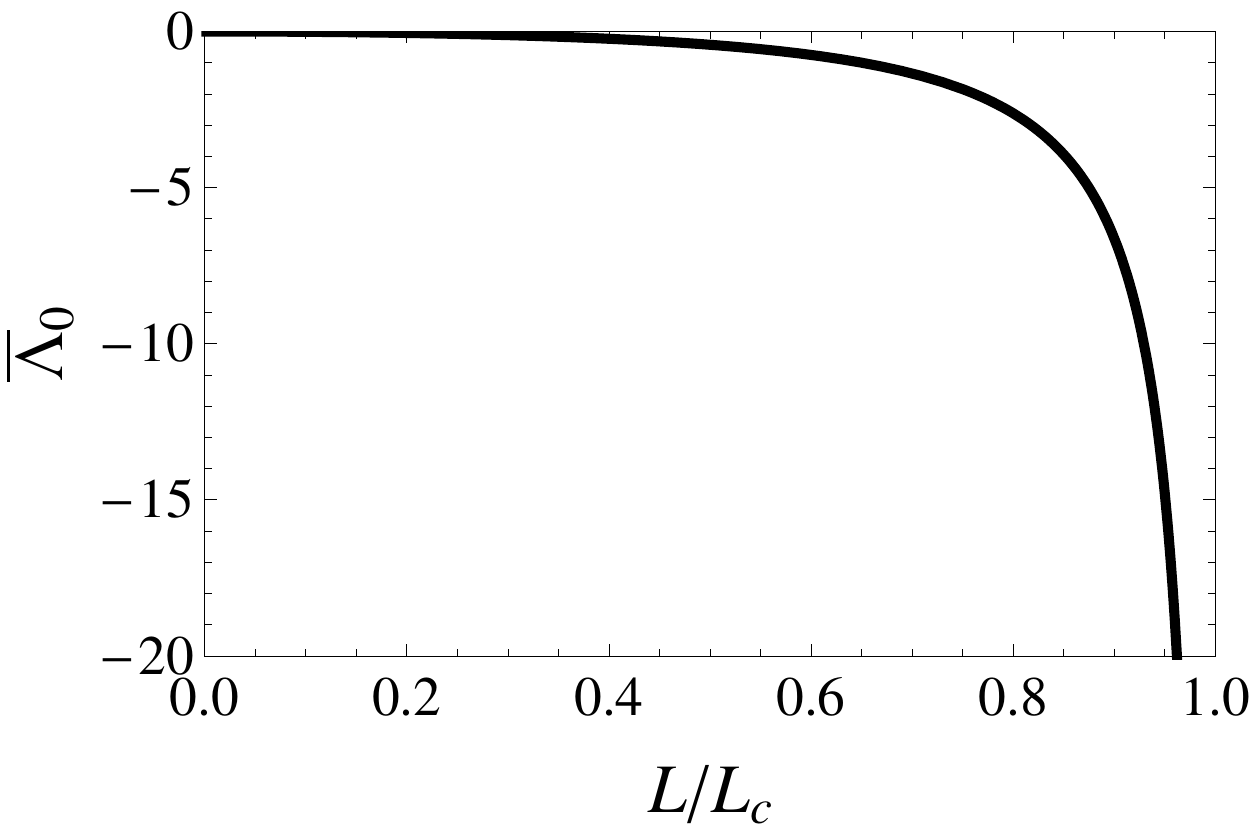}} &
\subfigure[]{\includegraphics[scale=0.33]{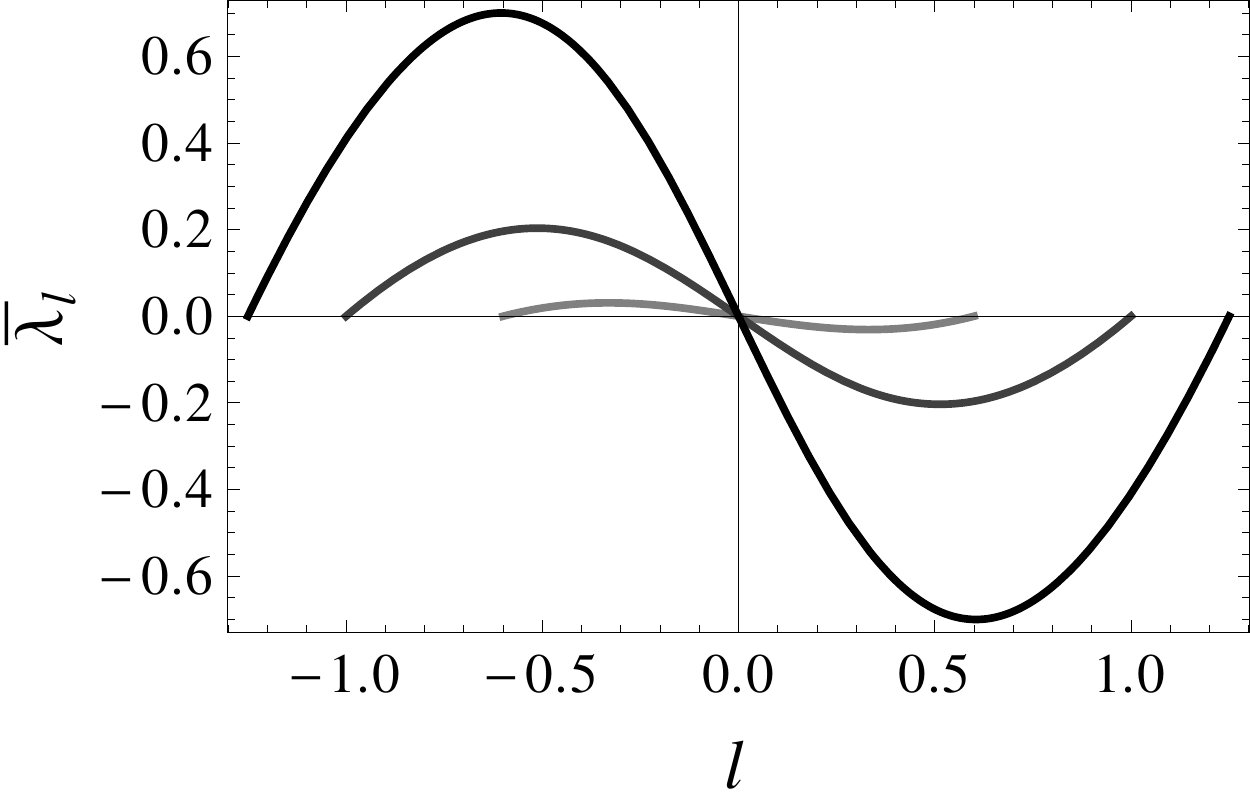}}
\end{tabular}
\caption{\label{fig1} \small (a) Catenoid between two identic rings. (b) $\Lambda$ as a function of arc-length
$l$ for values of boundary arc-length $L = 0.6, 1, 1.25$. The gray scale increases with $L$. (c)
Local extrema of the multiplier $\Lambda/(4\sigma R_0^2)$  as a function of the boundary arc-length
$L$. (d) $\lambda_l$ as a function of arc-length $l$ for the same values of $L$ as in (a).}
\end{center}
\end{figure}

It is curious that one never needed to identify the multiplier fields explicitly to isolate the  surface Euler-Lagrange equations.  If one were to stop here, however, would be a mistake:  for in the role that they play in quantifying the  forces necessary to constrain the tensor fields (\ref{GCM2D}), the multipliers also signal when surface
instabilities are present.\\
In this section, the partial differential equations describing these fields will be determined. For
simplicity,  examine the area ${\cal H}=\sigma$ with Euler-Lagrange equation $K=0$. In general, the
equation
\begin{equation} \label{TraceTab}
{\cal T}^{a}_{\phantom{a} a} =- \frac{1}{2 \sqrt{g}} \, {\cal L}_{\bf  \Lambda} \left(\sqrt{g} K\right)\,,
\end{equation}
follows by tracing over Eq.(\ref{TabdelgabKab}). Under the isometry ${\bf \Lambda}$,
(\ref{TraceTab}) implies that the mean curvature changes by a constant: ${\cal L}_{\bf  \Lambda}
K=-4 \sigma$. Combining this result with the contraction of Eq. (\ref{deltaKab}),
${\cal L}_{\bf  \Lambda} K = \left( -\nabla^2 + {\cal R} \right) \Lambda$, one obtains
\begin{equation} \label{n4}
\left(-\nabla^2 + {\cal R} \right) \Lambda  = -4 \sigma \,.
\end{equation}
The scalar $\Lambda$ is determined independently of the vector field $\lambda^a$. The differential
operator appearing here,  $\mathscr{L} = -\nabla^2 + {\cal R}$, also makes an appearance in the
second variation of area about any equilibrium geometry, which assumes the form $
\delta^2 A = \, \int dA \, \Phi \, \mathscr{L} \, \Phi$, where $\Phi$ is the normal deformation of the surface. The existence of negative eigenvalues signals a mode of instability of the surface.\\
As discussed elsewhere \cite{boundary} the appropriate boundary condition on $\Lambda$ in Eq. (\ref{n4}) is $\Lambda=0$. Its solution subject to this boundary condition is also unique.\\
To complete the determination of ${\bf \Lambda}$, note that the contraction of Eq. (\ref{deltagab})
implies that $\nabla_a \lambda^a=0$. The divergence of Eq. (\ref{deltagab}) then reads
\begin{equation} \label{n3}
(\nabla^2 + \frac{1}{2} {\cal R}) \lambda_a = - 2 K_{ab} \nabla^b \Lambda\,.
\end{equation}
A sufficient boundary condition is  $\lambda_a=0$. Given the function $\Lambda$,
the solution of  Eq. (\ref{n3}) is now unique. We will now show that the behavior of the multipliers
correlate with the stability of the equilibrium surface.\\
{\bf Example: Catenoid}. We will examine the behavior of the multipliers on a catenoid bounded by
two rings a fixed distance apart. Aligning the axis of symmetry along the $Z$ axis, its radius and
height, $R(l)$ and $Z(l)$, can be parameterizing in terms of arc-length $l$ along its meridians
(with $l=0$ on the neck of radius $R_0$, see Fig. \ref{fig1}(a)): $ R(l) = \, \sqrt{1 + l^2}$, $Z(l) =  \arcsinh \, l$, where all lengths are measured in units of $R_0$. The principal curvatures along the parallels and meridians are $C_\parallel = - C_\perp = 1 /R^2$. For simplicity, consider a  symmetric section of catenoid bounded by the parallel circles at $l=\pm L$, with corresponding radius $R_L$ and height $\pm Z_L$ respectively.\\
By symmetry $\Lambda$ is axially symmetric. Eq. (\ref{n4}) then assumes the form
\begin{equation} \label{LLambda}
\frac{\left(R \bar{\Lambda}' \right)'}{R} + 2 \, \frac{ \bar{\Lambda}}{R^4} = 1\,,
\end{equation}
where the prime indicates a derivative with respect to arc-length,
and $\bar{\Lambda} := \Lambda/(4 \sigma R_0^2 )$.
An exact solution of Eq. (\ref{LLambda}) exists. With the boundary conditions $\Lambda (\pm L) = 0$
it is given by
\begin{equation} \label{Lambdagen}
 \bar{\Lambda} = \frac{l}{4} \left(\frac{1}{R} \ln \left(R+ l\right) +  l\right) +
C_0 \left(\frac{l}{R} \ln \left(R+ l\right) - 1\right) \,,
\end{equation}
with integration constant,
\begin{equation}
C_0 = \frac{L}{4} \left(\frac{ L  R_L + \ln ( R_L +  L)} { R_L -  L \ln (  R_L + L)}\right)\,.
\end{equation}
Note that the global minimum, $\Lambda_0$, occurs at the neck where the curvature is highest.
Furthermore $\bar \Lambda_0 := \bar \Lambda(0)= -  C_0$  diverges as $L$ is increased to the value
$L_C  = 1.50888$ which occurs when $R_L = L  \ln (R_L + L)$ and the ratio of separation $h_L = 2
Z_L$ to diameter of the rings $D_L = 2 R_L$ is $h_L / D_L = 0.66274$. $\bar \Lambda$ is plotted as a
function of $l$ for several values of $L$ in the interval $[0,L_C]$ in Fig. \ref{fig1}(b). It is
negative everywhere in this interval.\\
The divergence of $\Lambda_0$ at $L = L_C$ correlates with the onset of an instability in
the catenoid as a minimal surface (see Fig. \ref{fig1}(c)). For let us expand $\Lambda$ in terms of
the eigenfunctions of the operator $\mathscr{L}$, $\Lambda= \sum_n C_n \Phi_n$, where $\mathscr{L}
\Phi_n = E_n \Phi_n$, so that Eq. (\ref{n4}) reads $\sum_n E_n C_n \Phi_n = -4\sigma$. Let $\Phi_0$
be the normalized ground state with eigenvalue $E_0$. Then
\begin{equation} \label{E0}
E_0 C_0  = -4 \sigma \int dA \Phi_0\,.
\end{equation}
If $\Phi_0$ is positive everywhere, the left-hand side of Eq. (\ref{E0}) is manifestly negative. If
$L$ is small, the catenoid approximates a cylinder with positive $E_0$. This implies that $C_0$ is
negative and thus so also is $\Lambda$, consistent with the exact solution. As $L\to L_C$, however,
$E_0\to 0$. At this value of $L$ Eq. (\ref{E0}) implies that $C_0$ must diverge, so that $\Lambda$
does also.  Thus an unexpected bonus of this framework is a reformulation of the analysis of
stability of minimal surfaces. $L= L_C$ is the maximum value of the meridian length for which the
catenoid is stable \cite{Fomenko}. Beyond $L = L_C$,  $E_0$ becomes negative and it can be shown
that $\Lambda$ changes sign.
Likewise, defining $\bar{\lambda}_a = \lambda_a/(4 \sigma R_0^2)$, Eqs. (\ref{n3}) are given by
\begin{equation}
\frac{\left(R \bar{\lambda}_l' \right)'}{R} - \frac{\bar{\lambda}_l}{R^4} = 2 \frac{\bar{\Lambda}'}{R^2} \,, \quad
\frac{\left(R \bar{\lambda}_\phi' \right)'}{R} - \frac{\bar{\lambda}_\phi}{R^4} = 0\,.
\end{equation}
Axial symmetry implies that the angular component vanishes:$\lambda_\phi=0$, consistent with the fact that the CM constraint equation along the polar direction vanishes identically: $C_\phi=0$. Thus, there is only a generalized force along the meridians.  The corresponding component $\lambda_l$ is plotted in Fig \ref{fig1}(d) for values of $l$ in the interval $[0, L_C]$. It is an antisymmetric function of $l$, possessing two extrema, one maximum and one minimum, vanishing at the neck where $l=0$.  Like $\Lambda$, $\lambda_l$ diverges at the onset of instability at $L=L_C$.

\section{Conclusions}

In this Letter it was shown how a surface can be treated as a Riemannian manifold endowed  with a metric that couples to a symmetric tensor field. The GCM equations impose a constraint on these two fields. No direct reference is made to the surface itself.\\
We have established a framework for studying surfaces that mimics gravity; the surface itself is an emergent entity. In the process, intriguing connections with a theory of metrics are revealed that are likely to be worth exploring.\\
The metric approach developed here--tweaked appropriately--is ideally adapted to study the recently proposed programmed swelling of thin polymer sheets \cite{Hanna}. This approach to interfaces and membranes has clear relevance to a number of problems in soft matter: fluctuations or membrane mediated interactions could be treated in a manner that sidesteps the difficulties of the height function representation. Numerically one could contemplate relaxing the GCM equations, but suppress violations in a controlled way by introducing large coupling constants.\\
One curiosity and unexpected virtue of this framework is that the derivation of the surface Euler-Lagrange equation never requires the explicit determination  of the Lagrange multipliers enforcing the constraints. These multipliers are, however, of considerable interest in their own right: it is they that quantify the strength of the coupling between the Riemannian metric and the symmetric tensor field shaping the manifold into a stationary state of the surface. A connection between conformal transformations and surface states has also emerged; its significance remains to be explored. More importantly, however, singularities in the multipliers correlate directly with instabilities in equilibrium surfaces. We have explored in some detail the behavior of these multipliers for surfaces minimizing area and, in particular, a soap film between two rings.\\
Extending this framework to higher dimensional surfaces or non-trivial backgrounds is not entirely
straightforward.  Unlike the two-dimensional case examined here, where the
contractions of the GCM constraints completely encode their geometrical content, these constraints
will need to be accommodated within the Hamiltonian in their full uncontracted glory. In
particular, the fortuitous similarity with the ADM formulation of general relativity encountered
here becomes an unreliable guidepost; the multiplier fields no longer assemble naturally into a
vector field.  What is more, the GCM equations will need to be supplemented with their Ricci
counterpart if higher codimensions are contemplated \cite{Spivak}.

\section*{Acknowledgements}

Support from DGAPA PAPIIT grant IN114510-3 and CONACyT grant 180901 is acknowledged.  We are also
grateful to Marcelo Dias and James Hanna for valuable comments.

\begin{appendix}

\section{Making connections and establishing contrasts} \label{appauxil}

It is useful to compare this approach with a variational framework introduced by one of the authors
several years ago which adopts a very different strategy \cite{auxil}.  In that approach $H$ is
again constructed using the metric and extrinsic curvature as independent variables. In contrast,
however, these variables are connected to the embedding functions through the Gauss-Weingarten
structure equations. One thus constructs  the Hamiltonian
\begin{eqnarray}
H_C &=& H [g_{ab}, K_{ab}] \\
&+& \int dA \,\left[ \mathbf{f}^a \cdot (\mathbf{e}_a-\partial_a\mathbf{X}) + \lambda_\perp^a \mathbf{e}_a\cdot \mathbf{n} + \lambda_n (\mathbf{n}^2 -1)  \right] \nonumber\\
 &+& \int dA \,\left[ \Lambda^{ab} (K_{ab}- \mathbf{e}_a\cdot \partial_b\mathbf{n}) +
\,\lambda^{ab}(g_{ab}- \,\mathbf{e}_a\cdot\mathbf{e}_b)\right] \,, \nonumber
\end{eqnarray}
implementing the definitions of $g_{ab}$ and $K_{ab}$ in terms of the tangent vectors
$\mathbf{e}_a$ and the normal $\mathbf{n}$, as well as the connection of the latter
to $\mathbf{X}$ by introducing appropriate Lagrange multipliers.  One is then free to treat each of
these variables independently. In particular, the translational invariance of $H$ implies the
existence of a conserved stress tensor.  In this framework, $\mathbf{X}$ only appears in the tangential
constraint so that
\begin{equation} \label{delHC}
\delta_{\bf  X} H_C = \int dA \nabla_a \mathbf{f}^a \cdot\delta\mathbf{X}\,,
\end{equation}
modulo a boundary term. Thus, in equilibrium, $\nabla_a \mathbf{f}^a=0$, or the stress $\mathbf{f}^a$ is
conserved. $\mathbf{f}^a$ is constructed using the remaining Euler-Lagrange equations. One finds
\cite{auxil}, $\mathbf{f}^a = f^{ab} \mathbf{e}_b + f^a \mathbf{n}$\,,
where
\begin{equation} \label{fabfa}
f^{ab}= T^{ab} -  H^{ac} K_c{}^b \,,\quad f^a = - \nabla_b  H^{ab}\,,
\end{equation}
and $T^{ab}$ and $ H^{ab}$ were defined in Eq. (\ref{TH}). It depends only on the geometry. In the new framework it is not obvious how to address the Euclidean invariance of the surface Hamiltonian, never mind the conservation laws that it implies, when the surface and its background do not yet exist.\\
The normal projection of the conservation law reads
\begin{equation} \label{nablafn}
\nabla_a f^{a} - K_{ab} f^{ab} =0\,;
\end{equation}
its tangential counterparts $\nabla_a f^{ab} + K^{b}_{\phantom{b}a} \, f^a =0$, are the statement of
reparametrization invariance. Notice that if ${\cal H}$ depends only on $g_{ab}$, Eq.
(\ref{nablafn}) reduces to the statement that $-  K_{ab}  T^{ab} =0$. This also  justifies  the
strategy that was adopted to identify the surface Euler-Lagrange equation.

\end{appendix}

\end{document}